\RequirePackage{fix-cm}
\RequirePackage{rotating}
\documentclass{svjour3}                     
\smartqed  
\usepackage{graphicx}
\usepackage{color}
\usepackage{natbib}
\usepackage{tipa}
\usepackage{amssymb}
\usepackage{amsmath}
\usepackage{faktor}
\usepackage{amsbsy}
\usepackage{bm}
\usepackage{txfonts}
\usepackage{tabularx}
\usepackage{marvosym}
\usepackage{overpic}
\usepackage{comment}
\usepackage{CJK}
\usepackage{enumerate}
\usepackage{xcolor}
\usepackage{booktabs}
\usepackage{multirow}
\usepackage{enumitem}
\usepackage{caption}
\usepackage{makecell}
\definecolor{blue}{rgb}{0,0,0}
\definecolor{red}{rgb}{0,0,0}

\graphicspath{{figure/}}
\title{Particle reconstruction of volumetric particle image velocimetry with the strategy of machine learning}
\titlerunning{PR of volumetric-PIV with strategy of ML}
\author{Qi Gao$^1$ \and Shaowu Pan$^2$\Letter \and Hongping Wang$^3$ \and Runjie Wei$^3$ \and Jinjun Wang$^4$ }
\institute{Q. Gao \at
              School of Aeronautics and Astronautics, Zhejiang University, Hangzhou, China, 310027
           \and
           S. Pan (\Letter) \at
           Department of Aerospace Engineering, University of Michigan, Ann Arbor, MI, 48105 \\
           \email{shawnpan@umich.edu}           
           \and
           H. P. Wang \and R. J. Wei \at
           MicroVec. Inc., Beijing, China, 100191
           \and
           J. J. Wang \at
           Key Laboratory of Fluid Mechanics, Ministry of Education, Beihang University, Beijing, China, 100191}
\date{Received: date / Accepted: date}
\begin{document}

\maketitle
\begin{abstract}
Three-dimensional particle reconstruction with limited two-dimensional projections is an under-determined inverse problem that the exact solution is often difficult to be obtained. In general, approximate solutions can be obtained by iterative optimization methods. In the current work, a practical particle reconstruction method based on a convolutional neural network (CNN) with geometry-informed features is proposed. The proposed technique can refine the particle reconstruction from a very coarse initial guess of particle distribution that generated by any traditional algebraic reconstruction technique (ART) based methods. Compared with available ART-based algorithms, the novel technique makes significant improvements in terms of reconstruction quality, {robustness to noises}, and at least an order of magnitude faster in the offline stage.
\end{abstract}

\label{intro}
Particle image velocimetry (PIV) is a widely used technique for measuring velocity fields \cite{wang2020effect,hong2020snow}. With volumetric PIV measurement, complex flows can be investigated regarding their three-dimensional three-component (3D3C) flow structures. Among all the 3D3C measurement methods, tomographic PIV (Tomo-PIV) proposed by \cite{Elsinga06} has been proved on its success of making an accurate measurement with fine spatial resolution under a fairly high particle seeding density of 0.05 ppp (particle per pixel). The key procedure of Tomo-PIV is the particle reconstruction (PR), which is a process of solving inverse projection problem from two-dimensional particle images to 3D intensity distribution of particles. In the original article of Tomo-PIV by \cite{Elsinga06}, the multiplicative algebraic reconstruction technique (MART) based on the maximum entropy criterion was introduced to reconstruct the 3D particle field. Since then, numerous advanced techniques have been developed to optimize the 3D particle reconstruction for improving either accuracy or efficiency, which has been well-reviewed by \cite{Scarano2012} and \cite{Gao2013}. Most available particle reconstruction techniques are based on MART algorithm, such as the spatial filtering MART (SF-MART), which applies spatial filtering on the reconstructed particle intensity field after each MART iteration \cite{Discetti2013}. SF-MART provides a better reconstruction quality than the traditional MART algorithm, which will be tested and compared with the new technique in the current work.

For the PR problem, with the increase of particle seeding density, the reconstruction quality decreases rapidly due to the issue of ghost particles, which is a fake particle unexpectedly generated at the intersections of light of sight (LOS). Many algorithms were proposed to \textcolor{black}{accelerate the optimization of PR by providing a good initialization}: \cite{Worth08} used multiplicative first guess (MFG) as a precursor to the standard MART approach, which can provide a reasonably accurate solution as the initial condition for MART iteration and also accelerate the convergence. \cite{Atkinson09} further proposed a multiplicative LOS (MLOS) estimation to determine the possible particle locations without requiring the weighting matrix as MFG. Besides having a good initialization, the removal of ghost particles can substantially improve the reconstruction quality. The joint distribution of peak intensity and track length can be used to successfully separate ghost particles and actual particles in certain cases \cite[]{Elsinga14}. A simulacrum matching-based reconstruction enhancement (SMRE) technique proposed by \cite{Silva13} utilizes the characteristic shape and size of actual particles to remove ghost particles in the reconstructed intensity field. The Shake-The-Box (STB) approach \cite{Schanz14, Schanz16} estimates trajectories based on previous time steps. The particle locations are consequently corrected by the Iterative Reconstruction of Volumetric Particle Distribution (IPR) proposed by \cite{Wieneke13}. STB has a considerable improvement compared to MART in both accuracy and particle concentration. For time-resolved image acquisition, sequential motion tracking enhancement MART (SMTE-MART) proposed by \cite{Lynch15} also produces a time-marching estimation of the object intensity field based on an enhanced guess, which is built upon the object reconstructed at the previous time instant. This method yields superior reconstruction quality and higher velocity field measurement precision when compared with both MART and MTE-MART \cite{Novara2010}. For single volume reconstruction, some new reconstruction schemes were developed. Intensity-enhanced MART (IntE-MART) uses a histogram-based intensity reduction to suppress the intensity of ghosts \cite{WangHP2016}. \cite{Gesemann10} solved the volume intensity using an optimization algorithm based on constrained least squares strategies and L1-regularization. \cite{Ye15} proposed a dual-basis pursuit approach for particle reconstruction, which yielded higher reconstruction quality comparing with MART in 2D simulations. In order to reduce the computational time, \cite{bajpayee2017} presented a memory-efficient and highly parallelizable method based on a homography fit synthetic aperture refocusing method. Rather than a `voxel-oriented' approach, \cite{BenSalah2018} proposed an `object-oriented' approach called Iterative Object Detection-Object Volume Reconstruction based on Marked Point Process (IOD-OVRMPP) for the reconstruction of a population of 3D objects. The particle position can be directly obtained using this method.

With the development of machine learning in the field of image processing, designing a model based on machine learning to deal with various image-related tasks has become a hot topic. In the past few years, neural networks have been applied to particle image velocimetry. Machine learning has been utilized to replace traditional cross-correlation for velocity deduction with dense particle seeding \cite{Cai2019EXIF, Cai2019IEEE}. Recently, a series of work has been presented in a conference, `13th International Symposium on Particle Image Velocimetry' (ISPIV 2019, Munich, Germany, July 22-24). For example, \cite{Lagemann2019} applied convolutional neural networks (CNN) to PIV and achieved similar effects of traditional cross-correlation algorithms. Liang et al. \cite{liang2020filtering} used CNN as a filtering step after several MART iterations for particle reconstruction. However, at the moment, most existing works on applying machine learning to PIV are two dimensional while an investigation of applying machine learning on particle reconstruction, as a fully three-dimensional application, is still lacking. {In this work, we present a novel machine learning framework (`AI-PR') using CNN~\cite{lecun1995convolutional} for 3D particle reconstruction problem.}

{This paper is organized as follows. In Section~2.1, the mathematical formulation of particle reconstruction is presented. In Section~2.2--2.3, the proposed architecture of AI-PR is described. In Section~3, as a preliminary study,  comparison of AI-PR against traditional SF-MART based algorithms on synthetic point cloud data is presented in terms of reconstruction quality, computational efficiency and robustness to noises. Finally, conclusions and future directions are summarized in Section~4.}

\section{Principle of particle reconstruction with machine learning}

\subsection{Particle reconstruction in TPIV as an inverse problem}

\textcolor{blue}{Since we cannot directly measure the 3D discrete particle field, we consider recovering the continuous 3D light intensity distribution resulting from the scattering by particles \cite{Elsinga06} from several 2D projections as an inverse problem \cite{minerbo1979ment}. For simplicity, we refer such inverse problem as particle reconstruction.}  Consider a fixed three dimensional orthogonal coordinate system, $(x,y,z) \in \mathbb{R}^3$, the unknown \textcolor{blue}{light intensity} field can be viewed as a continuous source function $f \in C^0(\mathcal{D})$ satisfying,
\begin{equation}
    f(x, y, z ) \ge 0, \quad \iiint_{\mathcal{D}} f(x,y,z)dxdydz = 1,
\end{equation}
where $\mathcal{D} \subset \mathbb{R}^3$ is a compact support of $f$.

Assuming parallel projection (or point spread function), without loss of generality, a view can be defined as a rotation of coordinate system with respect to some certain origin. One can further introduce different translations for cameras but it is ignored in the context for better illustration. The coordinate in the rotated system is $(x',y',z') \in \mathbb{R}^3$ where $x'-y'$ plane is parallel to the projection plane of the view, i.e., $z'$ is parallel to the line of sight, determined by the following relation,
\begin{equation}
    \begin{bmatrix}x' & y' & z'\end{bmatrix}^\top = \mathbf{T} \begin{bmatrix}x & y & z\end{bmatrix}^\top,
\end{equation}
where the rotation matrix $\mathbf{T}$ specified by three Euler angles $\alpha,\beta,\gamma$ is defined as,
\begin{equation}
    \mathbf{T} =
    \begin{bmatrix}
    \cos \alpha \cos \gamma - \sin \alpha \cos \beta \sin \gamma && -\cos \alpha \sin \gamma - \sin \alpha \cos \beta \cos \gamma && \sin \alpha \sin \beta \\
    \sin \alpha \cos \gamma - \cos \alpha \cos \beta \sin \gamma && -\sin \alpha \sin \gamma + \cos \alpha \cos \beta \cos \gamma && -\cos \alpha \sin \beta \\
    \sin \beta \sin \gamma && \sin\beta \cos \gamma && \cos \beta.
    \end{bmatrix}
\end{equation}

In practice, there are $J$ views, i.e., the number of cameras, usually ranging from 4 to 6. For each $j$-th view, the two dimensional projection field $g_j(x',y')$ is given as,
\begin{equation}
    g_j(x',y') = \int_{-\infty}^{\infty} f\left(\mathbf{T}^{-1}_j \begin{bmatrix}x' & y' & z'\end{bmatrix}^\top\right) dz'.
\end{equation}

\begin{figure}[htbp]
\centering
\includegraphics[width=0.7\textwidth]{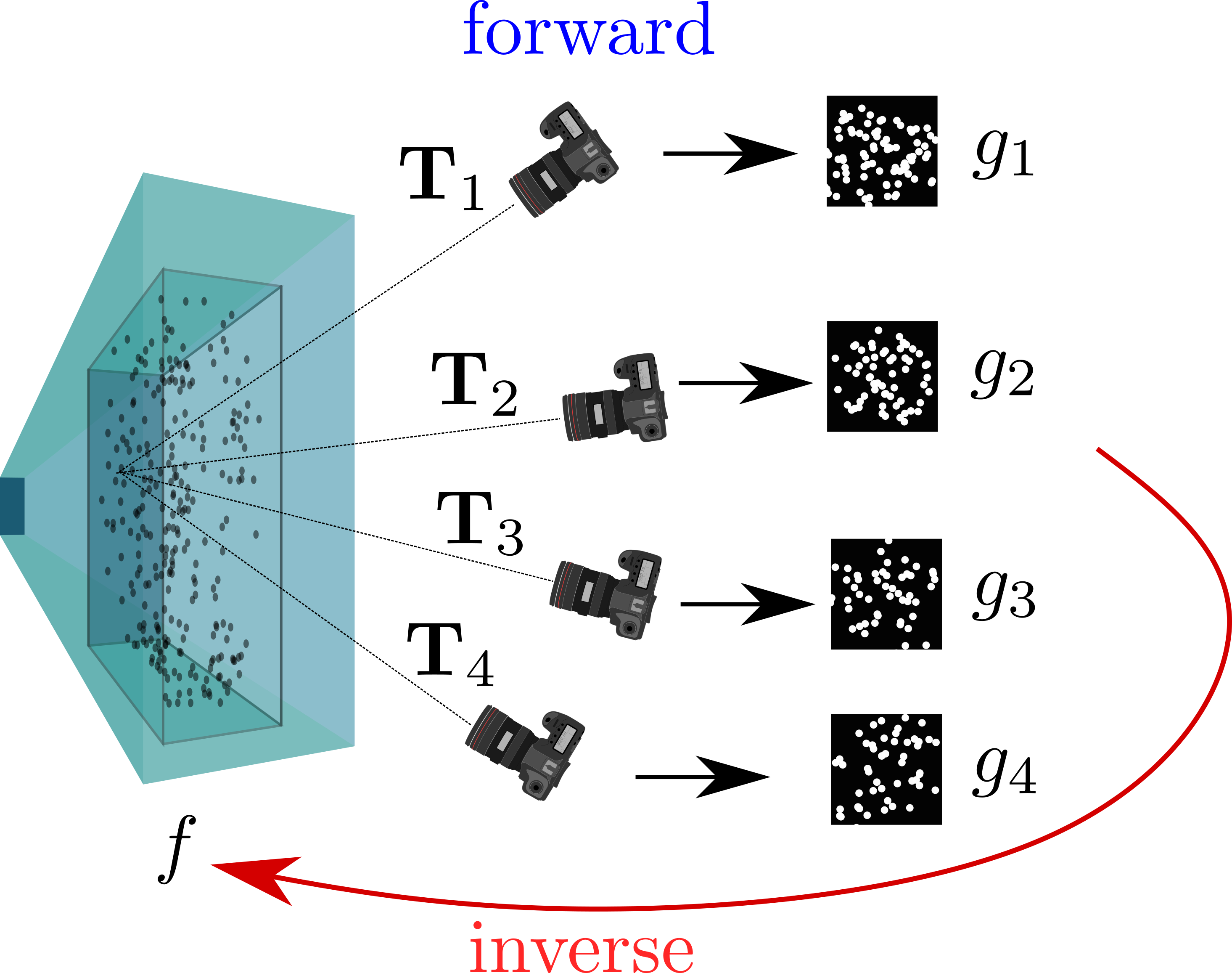}
\caption{Illustration of particle reconstruction as an inverse problem.}
\label{fig:0}
\end{figure}

As illustrated in Fig.~\ref{fig:0}, the goal of the inverse problem is to find the source function $f$, given projection data $\{g_j\}_{j=1}^{J}$ in the discretized form, i.e., $f$ are pixelized as function dealing with 3D matrix and $g_j$ as 2D images. Unfortunately, it is known to have infinite number of solutions satisfying all the above conditions \cite{guenther1974reconstruction, huesman1977effects}. Most often, additional conditions, e.g., entropy maximization~\cite{minerbo1979ment}, is considered to enforce uniqueness.

\subsection{Learning particle reconstruction field via CNN}

\subsubsection{CNN as a general and powerful field processing tool}

\begin{figure}[htbp]
\centerline{\includegraphics[width=0.9\textwidth]{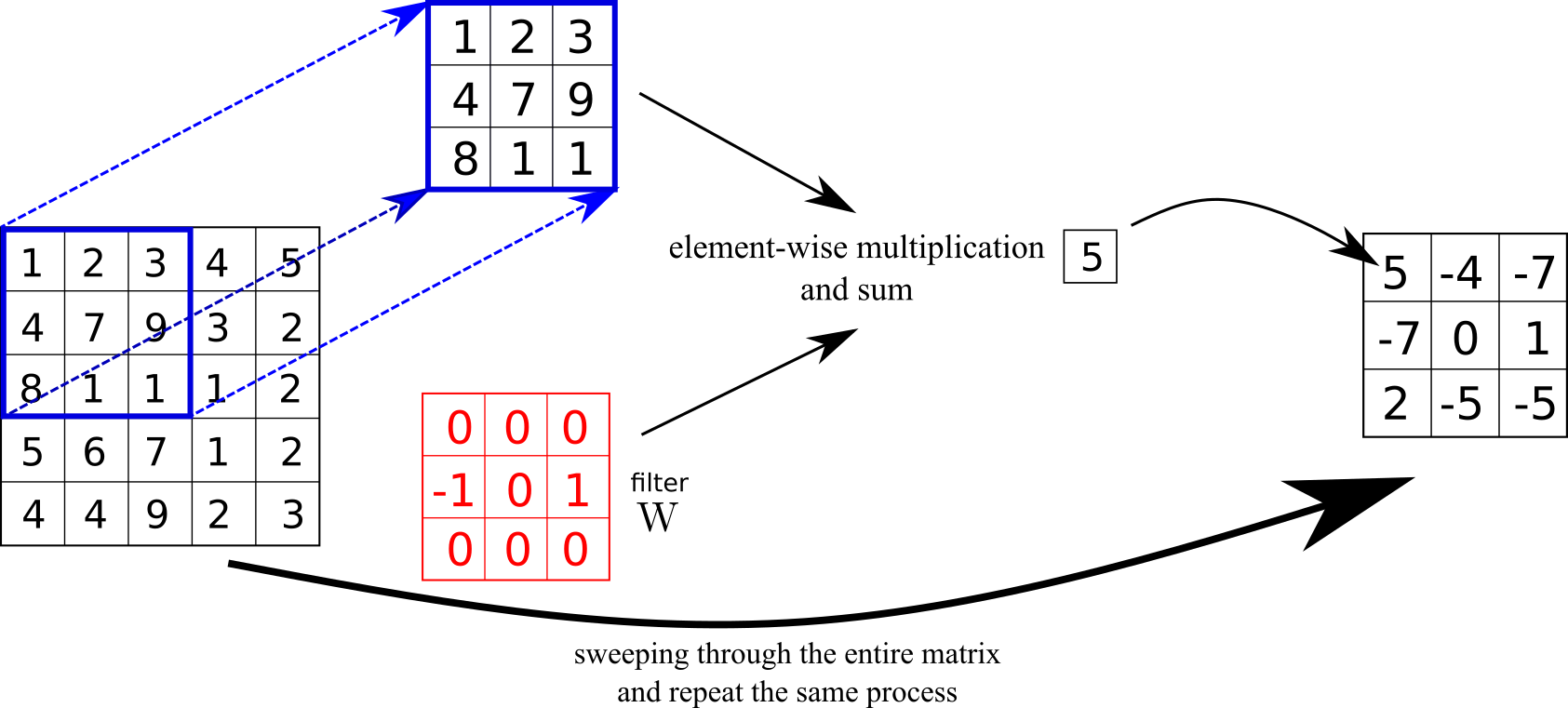}}
\caption{Illustration of 2D convolution of $3\times 3$ kernel $W$ on a $5 \times 5$ matrix.}
\label{fig:00}
\end{figure}

In recent years, with the increasing amount of data and computational power, CNN has become quite popular in many science and engineering communities with remarkable performance against traditional methods. Several examples include image processing: classification~\cite{wang2016cnn,krizhevsky2012imagenet}, object recognition~\cite{liang2015recurrent}, segmentation~\cite{milletari2016v}, inverse problem~\cite{mccann2017convolutional}; prediction of aerodynamics~\cite{bhatnagar2019prediction,guo2016convolutional}; model-order-reduction of fluid flows~\cite{lee2019model}. The main idea is to process the field with \textcolor{blue}{\emph{convolution} with non-linear activation} that leverages the locality (translation-equivariance) of the solution for many problems involving mapping on the spatial field, e.g., computing spatial derivative or average of a nonlinear function of the field. In Fig.~\ref{fig:00} we give an example of 2D \emph{linear} convolution on images. Note that by performing the convolution operation, the original $5\times 5$ image is transformed into a $3\times 3$ image. Such shrinkage in the image size is not favored in the context of deep CNN~\cite{goodfellow2016deep} since one would prefer the size of the output of CNN to be the same as the input size, which is the case in particle reconstruction problem. In this case, we consider padding zeros~\cite{goodfellow2016deep} around the original images so that the convoluted image would have the same size as the original image. Further, one can apply element-wise operation of an known nonlinear activation function $\sigma(\cdot):  \mathbb{R} \mapsto \mathbb{R}$ to make convolution nonlinear. Typical activation functions can be tanh, ReLU~\cite{goodfellow2016deep}, etc. In this work, we use ReLU activation function that is defined as,
\begin{equation}
\label{eq:relu}
\sigma_{\textrm{ReLU}}(x) = \textrm{max}(0,x).
\end{equation}
For more information about CNN, interested readers are referred to the following excellent reviews~\cite{rawat2017deep,mccann2017convolutional,aloysius2017review,zhiqiang2017review}.

\subsubsection{Mathematical formulation of a single 3D convolutional layer}

Recall that we are interested in applying the 3D analogy of 2D convolution introduced in the previous section for particle reconstruction. Instead of an image, the input of 3D CNN is a 3D Cartesian field or a 3D tensor. Note that from the previous section, filter $W$ uniquely determines the convolution. Therefore, it is straightforward to see that if one performs convolution on the same original image with $Q$ different filters, one can end up with $Q$ output images, which denoted as a 4D tensor. In the community of image processing, a single image at each layer is called a channel, which comes from the RGB channels in digital images~\cite{baxes1994digital}. Thus from now on, we denote the shape of a general 3D multi-channel field as a 4D tensor with shape $N_x \times N_y \times N_z \times Q$, with $N_x, N_y, N_z$ representing the size of the 3D Cartesian field and $Q$ representing the number of channels. For convenience, the ensemble of $Q$ filters is called \emph{kernel}, which uniquely determines the above $Q$ convolutions from a 3D tensor to a 4D tensor. This concept can be generalized to convolutions between any 3D multi-channel field with different numbers of channels.

Now we consider the $s$-strided convolution operation with the generalized kernel $\mathbf{K} \in \mathbb{R}^{L\times M \times N \times Q \times Q^{'}}$ on 3D $Q$-channel field $\mathbf{V} \in \mathbb{R}^{N_x \times N_y \times N_z \times Q}$ with an output as 3D $Q^{'}$-channel field $\mathbf{Z} \in \mathbb{R}^{N^{'}_x \times N^{'}_y \times N^{'}_z \times Q^{'} }$ as $\mathbf{Z} = c(\mathbf{K}, \mathbf{V}, s)$. \textcolor{blue}{$L,M,N$ are positive odd numbers representing the width of the 3D convolutional filter $\mathbf{K}$ in each direction. $s$ is the stride length in convolution.} Additionally, this kernel $\mathbf{K}$ contains $Q\times Q^{'}$ such filters defined in Section~2.2.1. Specifically, for index $1 \le i \le N^{'}_x$, $1 \le j \le N^{'}_y$, $1 \le k \le N^{'}_z$, and channel index $1 \le q' \le Q'$, combining with zero-padding in Eq.~\ref{eq:zeropadding} to avoid shrinkage of image size so as to enable deeper neural networks,
\begin{equation}
    \label{eq:zeropadding}
    \mathbf{\tilde{V}}(l,m,n;i,j,k) =
    \left\{ \begin{array}{ll}
      \mathbf{V}_{(i-1)s+l,(j-1)s+m, (k-1)s+n, q } & \textrm{if } 1 \le (i-1)s+l \le N_x  \\& \textrm{and } 1 \le (j-1)s+m \le N_y \\
      & \textrm{and } 1 \le (k-1)s+n \le N_z \\
      0 & \textrm{otherwise}\\
\end{array} \right.,
\end{equation}
we have the following general expression for zero-padding convolution operation,
\begin{align}
    \label{eq:conv}
    \mathbf{Z}_{i,j,k,q'} &= c(\mathbf{K}, \mathbf{V}, s)_{i,j,k,q'} \\
    & = \sum_{q=1}^{Q} \sum_{l=\frac{1-L}{2}}^{\frac{L+1}{2}}\sum_{m=\frac{1-M}{2}}^{\frac{M+1}{2}}\sum_{n=\frac{1-N}{2}}^{\frac{N+1}{2}} \mathbf{\tilde{V}}(l,m,n;i,j,k) \mathbf{K}_{l,m,n,q,q'},
\end{align}
\textcolor{blue}{where $l,m,n$ are indices of 3D filters, e.g., $-1\le l,m,n \le 1$ for $L=M=N=3$. It is well-known that convolution operation has a close connection to finite difference. For example, when $L=3$, such convolution operation contains finite difference approximation of first and second order spatial derivatives.}

After obtaining the output field $\mathbf{Z}$, an element-wise nonlinear activation function $\sigma(\cdot):  \mathbb{R} \mapsto \mathbb{R}$ is applied on $\mathbf{Z}$. Finally, the whole process including the nonlinear activation above defined in Eq.~\ref{eq:conv_layer} is called a \emph{convolutional layer} $\mathcal{C}$ without pooling,
\begin{equation}
\label{eq:conv_layer}
    \mathbf{V}' = \sigma(\mathbf{Z}) = \sigma(c(\mathbf{K}, \mathbf{V}, s)) = \mathcal{C}(\mathbf{V}) \in \mathbb{R}^{N^{'}_x \times N^{'}_y \times N^{'}_z \times Q^{'} }.
\end{equation}

In summary, the above nonlinear convolution transforms a 3D $Q$-channel tensor into another 3D $Q^{\prime}$-channel tensor. It is important to note that, to fully determine such convolution, one just needs to determine the filters in the kernel, which will be discussed in Section 2.3.

\subsubsection{Geometry-informed input features}

Instead of naively taking input as $J$ images from the cameras, we consider input for the 3D CNN as the particle field generated by MLOS method: $\mathbf{E}_{\textrm{MLOS}}$ in Eq.~\ref{eq:mlos}. Because the geometrical optics information, i.e., directions and positions of all the cameras, are naturally embedded, $\mathbf{E}_{\textrm{MLOS}}$ is geometry-informed. 
\begin{equation}
\label{eq:mlos}
    \mathbf{E}_{\textrm{MLOS}}(x,y,z) = \prod_{j=1}^{J} g_j(\mathbf{\tilde{T}} \begin{bmatrix} x & y & z \end{bmatrix}^\top ),
\end{equation}
where $\mathbf{\tilde{T}}$ is the first two rows of $\mathbf{T}$.

\subsection{Architecture of AI-PR}

Unlike traditional MART-based methods which don't require any data, the framework of AI-PR depends on \emph{data}: a particle field $f$ and the corresponding 2D images projected on the cameras. However, in real experiments, it is often impossible to obtain $f$, i.e., the exact locations and intensity distribution of particles from a measurement. Hence, synthetic random particle field with resolution as $256 \times 256\times 128$ is employed as training and testing data. The synthetic particle fields and their images are generated following a typical way that has been widely used for testing PR algorithms. Details can be found in \cite{WangHP2016} and \cite{Ye15}. Four projections of particle fields were calculated from given mapping functions to simulate camera imaging. The initial MLOS field was then computed and prepared as input for the aforementioned 3D CNN.

\subsubsection{Overcoming memory bottleneck with a divide-and-conquer approach}

A key difference between 3D PR and most 2D computer vision problems is that \textcolor{blue}{a typical} 3D PR usually requires \textcolor{blue}{large magnitude of memory usage due to increase from $O(n^2)$ to $O(n^3)$, where $n$ roughly represents the resolution in one direction. For typical fluid dynamics problem, $n \sim O(10^2)$.} While convolutions are highly parallel and optimized on a typical graphical card which often contains limited memory. Then it becomes challenging to perform even \textcolor{blue}{mini-batch training \cite{goodfellow2016deep}} with 3D convolution operation on such huge 4D tensor on GPU especially in our framework where size-reduction operation, i.e., pooling, is absent. One of the direct solutions is to implement parallel 3D convolutions~\cite{gonda2018parallel, jin2018spatially}. Instead, we consider a divide-and-conquer approach. We divide the input MLOS field ($256\times  256\times 128$) into $4\times 4\times 4=64$ sub-fields ($64\times 64\times 32$) by dividing its length along each direction into 4 equal pieces. Then we assume that $f$ within each sub-field can be inferred from the $\mathbf{E}_{\textrm{MLOS}}$ in that subfield, \textcolor{blue}{i.e., assuming the mapping from MLOS field to the actual particle field can be approximated by an affordable non-local mapping, instead of a global mapping.} Effectively, the size of input and output for 3D CNN is reduced by a factor of $4^3=64$ while the number of data is increased by a factor of 64. \textcolor{blue}{Therefore, such divide-and-conquer approach makes the training for 3D CNN affordable while increases the number of data for mini-batch training.} Finally, we concatenate 64 output sub-field from CNN into the field with the same size as original ($256\times 256\times 128$). \textcolor{blue}{It should be note that when we test the model, we have to divide the corresponding testing dataset into several 3D blocks with the shape as $64\times 64 \times 32$, potentially with some overlapping. In the following section, we will show that we can apply our trained model on MLOS field of an even larger resolution thanks to the divide-and-conquer approach. But again, one must first divide the field into sub-field with shape of $64\times 64\times 32$ in order to apply the trained model.}

\subsubsection{Learning particle reconstruction with 3D CNN}

\begin{figure}[htbp]
\centerline{\includegraphics[width=\textwidth]{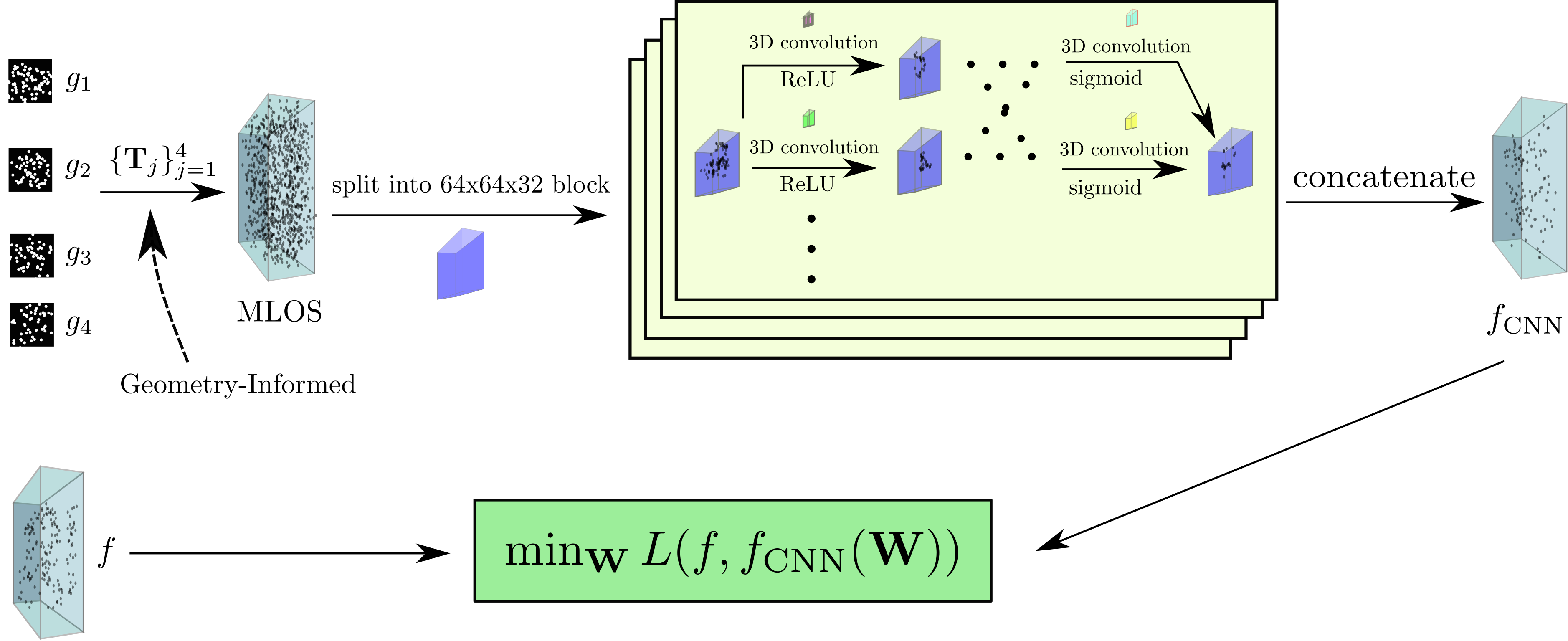}}
\caption{Schematic diagram of AI-PR.} 
\label{fig:1}
\end{figure}

Given the definition of a \emph{convolutional layer} in Section~2.2.2 as the building block, the architecture of AI-PR is illustrated in Fig. \ref{fig:1}. Following three major steps, one can obtain a good approximation of $f$ with 3D CNN output field $f_{\textrm{CNN}}$: 
\begin{enumerate}
\item $\mathbf{E}_{\textrm{MLOS}}$ is first calculated from multiple two-dimensional particle images by camera imaging, which is the same as traditional PR algorithms, while note that MLOS method has been noticed as a very good initial guess of particle field in MART-based algorithms. 
\item A 3D CNN is employed with batch normalization~\cite{Ioffe2015batch} such that it takes the input MLOS field $\mathbf{E}_{\textrm{MLOS}}$ and output $f_{\textrm{CNN}}$. 
\item Stochastic gradient-based optimization is performed, e.g., ADAM~\cite{kingma2014adam}, on kernel in all the layers to minimize the difference defined in Eq.~\ref{eq:loss} between \emph{known} $f$ from the training data and the corresponding $f_{\textrm{CNN}}$.
\end{enumerate}

\begin{equation}
\label{eq:loss}
L(f, f_{\textrm{CNN}}(\mathbf{W})) = \frac{ \sum_{j=1}^{M_{sub} M } \sum^{M_{res}}_{i=1} f^{(j)}(I_i, J_i, K_i) f^{(j)}_{\textrm{CNN}}(I_i, J_i, K_i; \mathbf{W})}{\epsilon + \sum_{j=1}^{M_{sub} M } \sum^{M_{res}}_{i=1} f^{(j)}(I_i, J_i, K_i)- f^{(j)}_{\textrm{CNN}}(I_i,J_i, K_i; \mathbf{W})},
\end{equation}
where superscript $(j)$ corresponds to $j$-th synthetic random particle field in the training data, $M$ is the total number of random clouds in the synthetic data, $M_{sub}$ is the number of sub-fields, e.g., 64, and $M_{res}$ is the total number of voxels of the sub-fields, e.g., $64\times 64 \times 32 = 131072$. $I_i, J_i, K_i$ are spatial indices of $i$-th voxel and $\mathbf{W}$ is the set of filters in all the kernels in the network.  $\epsilon$ is a small constant to avoid zero denominator. In addition, batch normalization~\cite{Ioffe2015batch} is used to accelerate the optimization.

Finally, after $f_{CNN}$ is trained (loss is sufficiently minimized over training data), to obtain a particle reconstruction for an unknown particle field $f$, one just need to compute MLOS field from the camera projections of $f$ and then take $\mathbf{E}_{\textrm{MLOS}}$ as input of the trained CNN to obtain $f_{\textrm{CNN}}$, which is supposed to be a good approximation of $f$.

\subsubsection{Structure of 3D CNN}

In this work, a deep 3D CNN with 12 hidden layers is employed. Size of input/output layer is $64\times 64 \times 32$ while that of hidden layers is $64\times 64 \times 32 \times 16$, i.e., each hidden layer has 16 channels with size unchanged. We consider ReLU activation function in Eq.~\ref{eq:relu} in each layer except the output layer where Sigmoid defined in Eq.~\ref{eq:sigmoid} is considered to ensure the output is bounded between 0 and 1. The convolution kernel of the input/output layer has size of $3\times 3 \times 3 \times 16$, while the other layers have size of kernel as $3 \times 3 \times 3 \times 16 \times 16$.

\begin{equation}
\label{eq:sigmoid}
\sigma_{\textrm{sigmoid}}(x) = \frac{1}{1+e^{-x}}.
\end{equation}

\subsubsection{Improving robustness with additive noise}

\textcolor{red}{Adding artificial noise to the synthetic data for assessment is a key issue of the evaluation. There are many types of noises such as white noise, Gaussian noise, Poisson noise and salt-and-pepper noise. White noise and salt $\&$ pepper noise are discrete signals whose samples are regarded as a sequence of serially uncorrelated random variables. Normally, this type of noise will not significantly affect PIV related algorithms, which can be easily reduced by applying pre-processing filter e.g. median filter. Poisson noise is commonly in weak light illumination during imaging, which is not a major noise under laser illumination of PIV measurement. 
On the other hand, Poisson noise can normally be approximated by Gaussian noise.
Hence, Gaussian noise is the most concerned type. Adding Gaussian noise to the dataset has been widely applied in many other seminal works \cite{cai2019particle,discetti2013spatial}.}
To investigate the robustness of AI-PR, we consider 20\% of the total $M$ training particle images biased with Gaussian noises. Different degrees of Gaussian noise is added to the four particle images. Following the typical way of adding noises \cite{WangHP2016, Cai2019EXIF}, the standard deviation $\sigma$ of the image noise is calculated with levels of $n\sigma$ for PR testing, where $n$ is from 0 to 0.2 with an interval of 0.05. It is noticed that the performance of the new algorithm is stable and accurate enough when the size of training data $M$ is over 500. 

\textcolor{red}{
However, it should be noted that calibration error also contributes significantly to the particle reconstruction. For the volumetric PIV calibration, one needs `self-calibration' \cite{wieneke2008volume} to significantly improve the accuracy of mapping functions and makes the uncertainty down to 0.1 pixels, which is small enough to guarantee the accuracy of particle reconstruction. In practice, this step is performed before we applying the new proposed particle reconstruction algorithm. Therefore, we believe that error from calibration is extremely reduced and is decoupled with the noise from the raw particle images. Comparing with the noise of particle images, the error of calibration after `self-calibration' is negligible. For this reason, we did not consider the negative effect of calibration error in this study, but only focused on the noise from particle images, which is an acceptable approach for studying particle reconstruction \cite{discetti2013spatial, WangHP2016}.}


\section{Results and discussions}

\subsection{Comparison setup}

In this section, we briefly describe the performance of AI-PR against the traditional SF-MART method \cite{Discetti2013} in terms of reconstruction quality, computational efficiency and robustness to noises. Again, recall that it is difficult to obtain the true particle field in a real experiment. Hence, as a preliminary study, the comparison between AI-PR and SF-MART method is conducted on synthetic random particle field data. The testing particle fields are generated in the same manner as the training set but with a different size of $780 \times 780 \times 140$. Note that since AI-PR is trained on the sub-fields anyway instead of the original field, we divide the $780 \times 780 \times 140$ into sub-fields with size $64 \times 64 \times 32$ with some overlap. Seeding density in the generation of random particle fields varies from ppp = 0.05 to 0.3 with an interval of 0.05. Noise level ranges from n = 0.05 to 0.3 with interval of 0.05. It is important to note that we only use the synthetic random particle field generated at ppp=0.2 with different noise levels for training AI-PR while the rest is for testing. Traditional PR methods: SF-MART with five and ten times iterations, together with proposed AI-PR and its input as MLOS field are considered for comparison against each other. The codes for the training and testing on AI-PR are developed with Tenserflow$^\text{TM}$ V1.13.1 \cite{TensorFlow} in Python (www.python.org) while MLOS and MART is developed with Matlab$^\circledR$ (MathWorks, Inc.). The computer used is an Intel x99 workstation with one CPU of E5-2696 V4, 64GB DDR4 memory and a RTX2080ti graphics processing unit.

\subsection{Comparison on cross-section of particle field}

\begin{figure}[htbp]
\centerline{\includegraphics[width=0.7\textwidth]{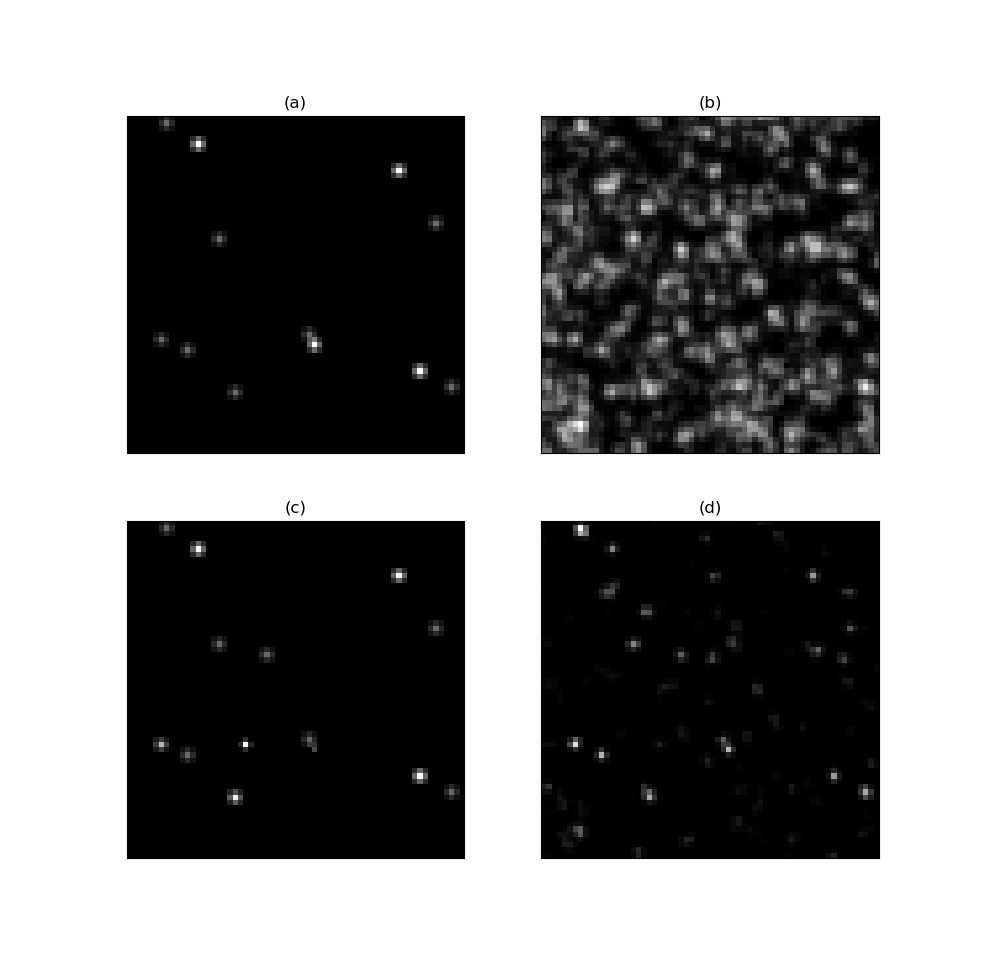}}
\caption{Cross-sections of particle field, (a) Synthetic field, (b)MLOS field, (c)AI-PR, (d) SF-MART field with 10 iterations.}
\label{fig:2}  
\end{figure}

Fig. \ref{fig:2} provides a central cross-section of a reconstructed particle field with ppp = 0.15 which is in the testing range. It is obvious that MLOS only gives a very coarse initial guess of potential particle location and intensity distribution, while AI-PR and SF-MART can recover better particle fields. Comparing further between AI-PR and MART methods, it is notable that SF-MART generates more ghost particles and has worse intensity distribution than AI-PR does. If the particle shape is looked closer, it can be found that MART-reconstructed particles have more ellipsoid shape, when AI-PR restores the spherical shape better. 

\subsection{Comparison on reconstruction quality, noise-robustness and computational efficiency}

\begin{figure}[htbp]
\centerline{\includegraphics[width=0.7\textwidth]{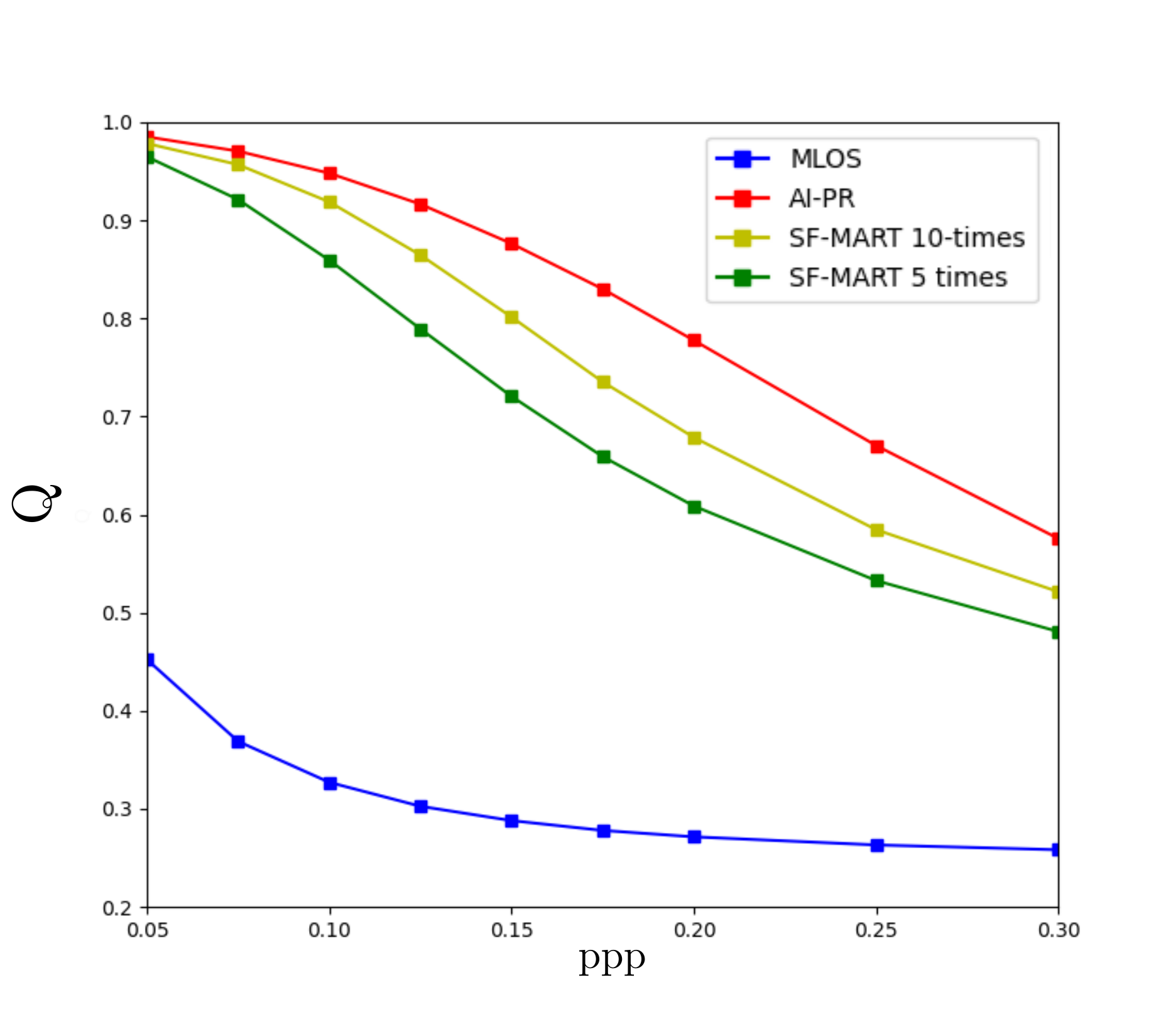}}
\caption{Quality factor $Q$ of different methods with varying seeding density \textcolor{blue}{from 0.05 to 0.3.}}
\label{fig:3a}
\end{figure}

\begin{figure}[htbp]
\centerline{\includegraphics[width=0.7\textwidth]{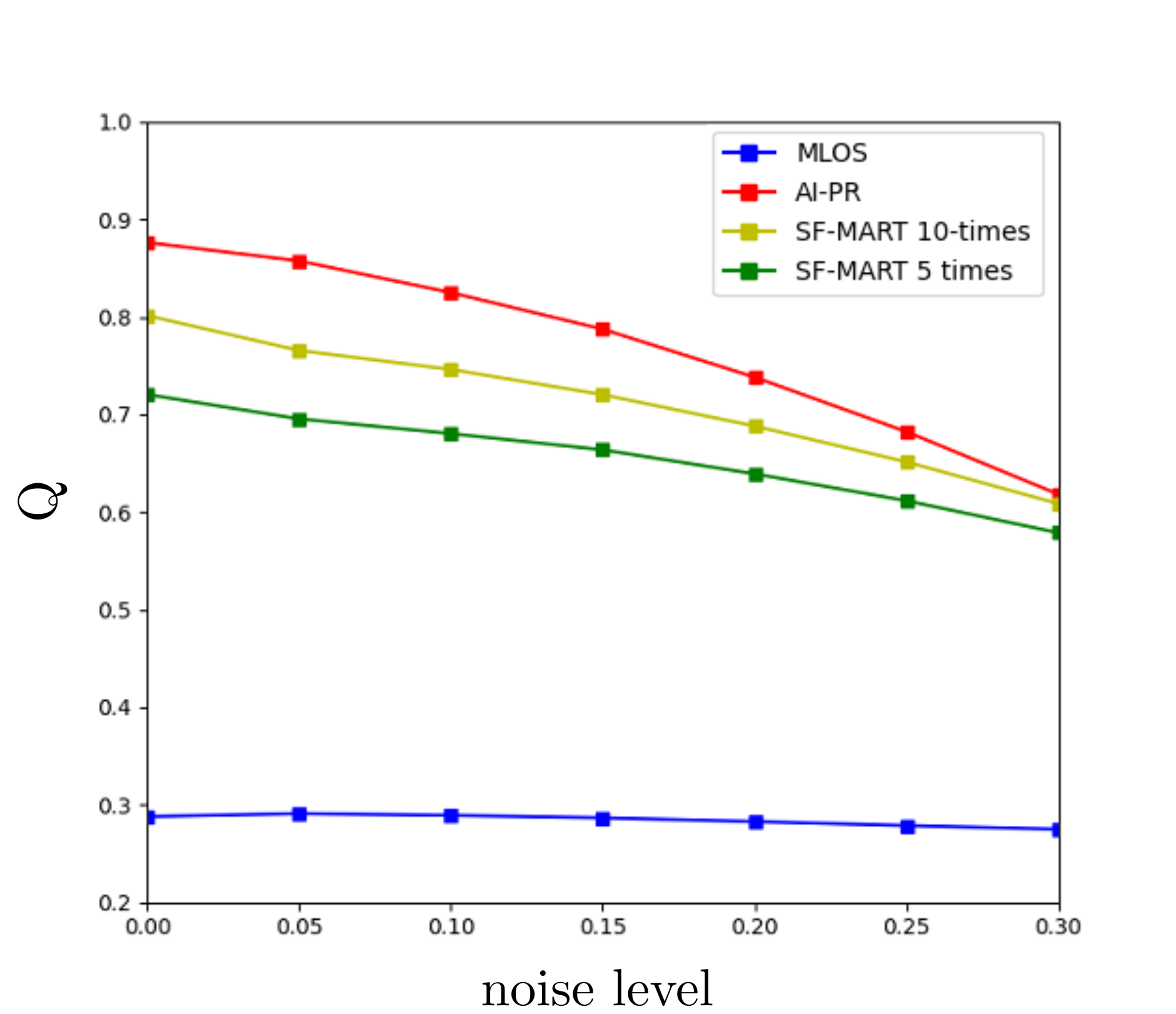}}
\caption{Quality factor $Q$ of different methods with varying noise levels \textcolor{blue}{from 0 to 0.3 with interval of 0.05 at ppp=0.15}}
\label{fig:3b}
\end{figure}

In terms of reconstruction quality, AI-PR shows its superiority to SF-MART methods as shown in Fig. \ref{fig:3a} and \ref{fig:3b}. The quality factor $Q$ defined in Eq.~\ref{eq:Q} is utilized for evaluating the accuracy and stability of the new technique, which is the correlation coefficient between the synthetic and reconstructed fields.

\begin{equation}
\label{eq:Q}
Q = \frac{ \sum^{M}_{i=1} f(I_i, J_i, K_i) f_{\textrm{CNN}} (I_i, J_i, K_i; \mathbf{W})}{\sqrt{ \sum^{M}_{i=1} f^2(I_i, J_i, K_i)} \sqrt{\sum_{i=1}^{M} f^2_{\textrm{CNN}}(I_i,J_i, K_i; \mathbf{W})}} ,
\end{equation}

In Fig. \ref{fig:3a}, all the methods are tested \textcolor{blue}{with varying particle density while} without noise. It is shown that AI-PR can recover the particle with significant improvements from MLOS field. Reconstruction quality $Q$ of AI-PR is much better than that of SF-MART methods. When ppp reaches 0.25, the $Q$ remains at around 0.7 for AI-PR, while SF-MART-10 reduces below 0.6. Next, the effect of noise is parameterized in Fig. \ref{fig:3b} at a fixed particle density ppp = 0.15, the $Q$ reduces with the increase of noise level for all methods, but AI-PR has the best stability against the biases.

%

\begin{table}[h!]
\centering
\caption{Summary of time cost for particle reconstruction.}
  \begin{tabular}{cc}
    \hline
    & time (sec)  \\ \hline
    MLOS & 512.5 \\
    SF-MART 5 iter. & 5333.5 \\
    SF-MART 10 iter. & 9881.5  \\
    AI-PR & \textbf{524.5}  \\ \hline
  \end{tabular}
  \label{tab:1}
\end{table}

As shown in Tab.~\ref{tab:1}, the algorithms of MLOS, SF-MART-5, SF-MART-10 and AI-PR takes wall-time as 512.5s, 5333.5s, 9881.5s, and 524.5s, respectively. 
Since AI-PR processing included the computing cost of MLOS and CNN, the actual computing time was only about 12s. However, it should be noted that although both SF-MART and the 3D CNN part in AI-PR shares the same computational complexity, leveraging the state of the art GPU computing power, the later can be highly parallel on GPU thus extremely fast while the former requires iterative calculation with dependency among voxels/pixels. Moreover, it is noticed that the training of CNN cost about 16 hours for 100 epochs and the SF-MART algorithm is not accelerated with GPU in the current work.

\section{Conclusions}

Robust and efficient 3D particle reconstruction for volumetric PIV has been a long standing problem in experimental fluid mechanics. Traditional SF-MART-based algorithms either suffer from expensive computational time or sensitivity to noises. As a preliminary study, the newly proposed AI-based technique shows its superior advantages of accuracy, efficiency (x10 faster), and robustness to noise on recovering particle locations and intensities from 2D particle images over traditional SF-MART-based algorithms. Overall, with its superior accuracy and robustness, we believe AI-PR technique is very promising to apply to more realistic experiments by increasing the training dataset. \textcolor{red}{However, as for the current work, the validation of our algorithm is limited to synthetic data rather than real experimental data. } Future work \textcolor{red}{should} focus on combining calibration of volumetric PIV with AI-PR training, and performing particle reconstruction directly from AI-PR without calibration and additional network training for different \textcolor{blue}{real} experimental cases.

\section*{Acknowledgements}
This work was supported by the National Key R \& D Program of China (No. 2020YFA040070), the National Natural Science Foundation of China (grant Nos. 11721202), the Program of State Key Laboratory of Marine Equipment (No. SKLMEA-K201910)

\section*{Competing interests}
The authors declare that they have no competing interests.

\section*{Authors' contributions}
Conceptualization, Methodology: Q. Gao, S. Pan, Runjie Wei. 
Software, Simulation: H. Wang. 
Writing original draft, Q. Gao.
Supervision, Funding acquisition: Q. Gao, J. Wang.
Reviewing and Editing: S. Pan. The authors read and approved the
final manuscript.

\bibliographystyle{spbasic}      
\bibliography{bmc_article}

\end{document}